\documentclass[graybox]{svmult}

\usepackage{bm}
\usepackage{amsfonts}
\usepackage{bbm}
\usepackage{bm}
\usepackage{amsmath}
\usepackage{amsmath}
\usepackage{mathptmx}       % selects Times Roman as basic font
\usepackage{helvet}         % selects Helvetica as sans-serif font
\usepackage{courier}        % selects Courier as typewriter font
\usepackage{type1cm}        % activate if the above 3 fonts are
\usepackage{makeidx}         % allows index generation
\usepackage{graphicx}        % standard LaTeX graphics tool
\usepackage{multicol}        % used for the two-column index
\usepackage[bottom]{footmisc}% places footnotes at page bottom
\newtheorem{OBS}{Remark}

\usepackage{hyperref}

\makeindex             % used for the subject index
\begin{document}

\title*{Dynamics of Cluster Synchronisation in Modular Networks: Implications for Structural 
and Functional Networks}
\titlerunning{Cluster Formation in Modular Networks} 
%for an abbreviated version of
\author{Jake Stroud,  Mauricio Barahona and Tiago Pereira}
\authorrunning{Stroud, Barahona, Pereira } 
%for an abbreviated version of
\institute{Jake Stroud \at University of Oxford, Oxford, UK, \\ \email{jake.stroud@wadh.ox.ac.uk}
\and Mauricio Barahona \at Department of Mathematics, Imperial College London, London, UK, \\ \email{m.barahona@imperial.ac.uk}
\and Tiago Pereira \at Department of Mathematics, Imperial College London, London, UK, \\ London Mathematical Laboratory, London, UK \\ \email{tiago.pereira@imperial.ac.uk}}
\maketitle

\abstract*{Experimental results often do not assess network structure; rather, the network structure is inferred by the dynamics of the nodes. From the dynamics of the nodes one then constructs a network of functional relations, termed the functional network. A fundamental question in the analysis of complex systems concerns the relation between functional and structural networks.  Using synchronisation as a paradigm for network functionality, we study the dynamics of cluster formation in functional networks. We show that the functional network can drastically differ from the structural network. We uncover the mechanism driving these bifurcations by obtaining necessary conditions for modular synchronisation.
}
\section{Introduction}
\label{sec:1}

When using methods from network science to study real-world complex systems, one is faced with the choice of constructing either a structural or functional network that describes the relationship between the interacting components of the system. Sometimes it is more achievable or desirable to measure the dynamics of components and posit that if two components display similar activity, they are in some way dynamically linked. This gives rise to data-driven \textit{functional networks}~\cite{bassett}. Conversely, in other situations, we may have access to \textit{structural networks} representing known physical links between components. The focus of much research in complex network theory is towards gaining a greater understanding of how functional and structural networks relate to each other~\cite{sporns1}.

To this end, much analysis has been conducted into network architecture and organisation. Recent results have shown that both functional and structural network representations of real world systems  typically display a modular architecture \cite{bassett2,bullmore,social,soc,cat,maur,Fortunato2010}. A network with a modular organization could be described as a network consisting predominantly of highly connected sub-graphs which have comparatively fewer connections to nodes outside the module. We are particularly interested in how functional modules relate to the structural modules of a network.

Synchronisation is a typical paradigm of dynamical network function. Such group collective behaviour appears with ubiquity in nature. Human hearts beat rhythmically because thousands of cells synchronise their activity \cite{cells}, and the collective behaviour of neurons in the brain has been shown to be linked to Parkinson's disease \cite{park} and epileptic seizures \cite{ep}. However, synchronisation does not have to be global, and can occur in particular subgroups or modules.

Over recent decades, synchronisation analysis has benefited from methods in the fields of graph theory and dynamical systems, and theories for global synchronisation have been established in terms of the network structure~\cite{msf, Heagy1995, Barahona2002, Arenas2008a}. Of particular interest is the stability of the synchronised state. If global synchronisation can be maintained, this amounts to a coherent state, while if the synchronised state becomes unstable it can serve to predict a transition in the organisation of the complex system \cite{synchbar, Belykh2004, Wu2005, August2011, msf, OClery2013}. 

In this chapter, we will study cluster synchronisation within modular networks of diffusively coupled oscillators. We predict the onset of stable and unstable module synchronisation by obtaining stability conditions within a densely connected module. Our model is based on the following assumptions:
\begin{svgraybox}
\begin{itemize}
\item[$\bullet$] The network has a modular structure such that nodes in one module have few connections to nodes outside the module.
\item[$\bullet$] Within modules, nodes have a high mean degree and share many common neighbours.
\end{itemize}
\end{svgraybox}

Under these assumptions, we can perform a stability analysis independently for each module, thereby avoiding a spectral decomposition of the network adjacency matrix which poses a significant challenge for large networks.  Typically, we would have to analyse as many equations as the number of nodes in a module, but because nodes within modules have a large number of common neighbours, we are able to reduce the analysis to a single equation describing the synchronisation of each module by bounding the dynamics of the external modules. Hence our analysis allows us to tackle the stability of each module independently, yet taking into account their influence on each other.

From our study of the local stability of synchronisation, we establish conditions for the persistence of stability under non-linear and linear perturbations. The stability of module synchronisation is determined as a function of: $(i)$ the module mean degree and matching index (defined in Section \ref{defn}), and $(ii)$ the isolated dynamics and coupling function. This analysis allows us to predict the formation and disintegration of functional modules depending upon the nature of the diffusive coupling between the components of the network. 

\newpage

Our results reveal that:  
\begin{svgraybox}
\begin{itemize}
\item[$\bullet$] The mean degree of the module dictates the onset of synchronisation.
\item[$\bullet$] Functional modules may not reflect fully the structural modules of a network.
\end{itemize}
\end{svgraybox}
As a consequence, the functional representation of a network can sometimes drastically differ from the underlying topological structure. Through the use of simulations we validate our analytical results and conclude with a discussion on how the functional network representation of a modular network relates to the underlying topological structure. The remainder of this chapter is organised as follows. Our model assumptions are formalised in Section \ref{model}. Both our analytical and numerical results are presented in Section \ref{results}. The derivation of our analytical results is then presented in Section \ref{derivation}. Finally we provide a conclusion and discussion in Section~\ref{summary}.

\section{The Model}\label{model}

In this section we formalise our model setup. Initially we formalise some basic graph definitions we make use of as well as describing the network class we intend to study. We then present the dynamical model describing the interaction between the components of the network. See Section \ref{notation} for remarks regarding notation we adopt throughout the chapter.

\subsection{Notation}\label{notation}

%We denote the Euclidean norm by
%\begin{equation*}
%\|\bm x\| := \sqrt[]{\bm x^{\top}\!\!\cdot \bm x}.
%\end{equation*}
The Jacobian matrix of a function $\bm f: \mathbb R^m \rightarrow \mathbb R^m$ at the point $\bm x$ is denoted by $D \bm f(\bm x)$. 

When discussing synchronisation in this chapter, we imply a $\delta$-synchronisation; that is, given two trajectories $\bm{x}(t)$ and $\bm{y}(t)$ we say that they are $\delta$-synchronised if the difference in their state vectors are within a neighbourhood of radius $\delta \ll 1$ 
at all large times $t$
$$ 
\| \bm{x}(t) - \bm{y}(t)  \| \le \delta \quad  \forall \, t > T(\delta). 
$$
The small parameter $\delta$ measures the quality of synchronisation, and depends on both the isolated dynamics and coupling function as well as the network structure.  This is particularly evident in the numerical simulations. To simplify the notation, we will omit the symbol $\delta$ when discussing $\delta$-synchronisation. 

We use the small `$o$' and big `$O$' notation to describe asymptotic behaviour. We write $f(x) = o(x)$ if $f(x)/x$ goes to zero as $x$ tends to infinity, and we write $f(x) = O(x)$ if $|f(x)/x|$ is bounded by a positive constant as $x$ tends to infinity.

\subsection{Graphs: Basic Definitions}\label{defn}
A graph $G$ is a set of $N$ nodes (or vertices) connected by a set of $E$ edges. Here, we will only consider simple, unweighted and undirected graphs; that is, graphs with no loops and where there is no order associated with the two vertices of each edge. We will also use the words `graph' and `network' interchangeably although a network commonly denotes a graph structure where some form of dynamics takes place on the nodes.

The adjacency matrix $\bm A$ encodes the topology of the graph, with $ A_{ij}=  1$ if $i$ and $j$ are connected and  $0$ otherwise. Clearly, $\bm A = \bm A^T$ for undirected graphs. The degree of node $i$ is the number of connections it receives, that is
\begin{equation*}
k_{i} = \sum_{j}^N A_{ij}.
\end{equation*}
The mean degree for a set of nodes $S$ with cardinality $|S| = n$  is then:
\begin{equation*}
{\langle k \rangle}_{S} = \frac{1}{n}\sum_{j \in S} k_j.
\end{equation*}

\begin{figure}[h]
\centering
%\begin{minipage}{1\textwidth}
\includegraphics[width=\textwidth]{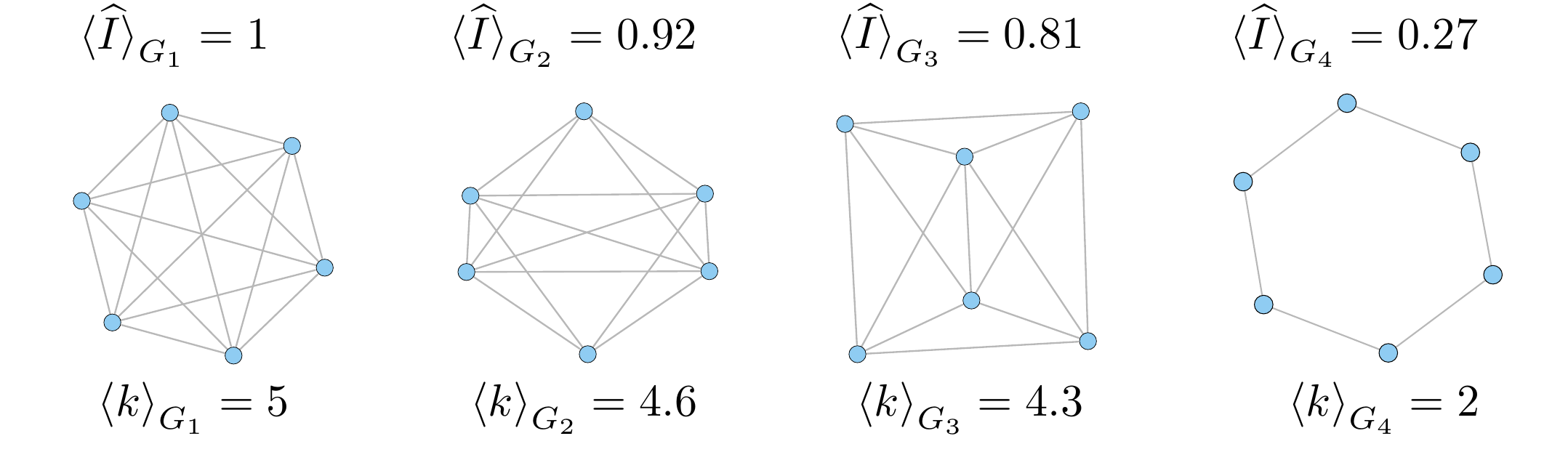}
\caption{Here we display four graphs, $G_1$, $G_2$, $G_3$ and $G_4$ with decreasing matching index and mean degree.}
\label{fig:graphs}
\end{figure}

%\newpage
We now define the \textit{matching index} of a graph~\cite{cat}, which will play an important part in our analysis. The neighbourhood of node $i$ is the set of nodes it shares an edge with: $\Gamma (i) = \{j| A_{ij} = 1\}$. Clearly, for simple graphs $|\Gamma(i)| = k_{i}$. The matching index of nodes $i$ and $l$ is the overlap of their neighbourhoods:
\begin{equation*}
I_{il} = |\Gamma(i) \cap \Gamma(l)| = {A_{il} + \sum_{n,m = 1}^{N} A_{in}A_{ml}} = (A + A^2)_{il}.
\end{equation*}
The normalised matching index is then:
\begin{equation}
\label{eq:match}
\widehat{I}_{il} = \frac{|\Gamma(i) \cap \Gamma(l)|}{|\Gamma(i) \cup \Gamma(l)|} = 
%\frac{A_{il} + \sum_{n,m = 1}^{N} A_{in}A_{ml}}{ k_{i} + k_{l} - A_{il} - \sum_{n,m = 1}^{N} A_{in}A_{ml}} 
\frac{|\Gamma(i) \cap \Gamma(l)|}{|\Gamma(i)| + |\Gamma(l)| - |\Gamma(i) \cap \Gamma(l)| } = 
= \frac{(A + A^2)_{il}}{k_{i} + k_{l} - (A + A^2)_{il}}.
\end{equation}
It follows that $\widehat{I}_{il} = 1$ if and only if $i$ and $l$ are connected to exactly the same nodes, i.e., $\Gamma(i) = \Gamma(l)$; whereas $\widehat{I}_{il} = 0$ if nodes $i$ and $l$ have no common neighbours~\cite{cat}. The mean matching index for a set of nodes $S$ with $|S| = n$ is then:
\begin{equation*}
{\langle \widehat{I} \rangle}_S = \frac{1}{n(n-1)} \sum_{\substack{i,j \in S \\ i \neq j}}\widehat{I}_{ij}.
\end{equation*}
Figure \ref{fig:graphs} shows graphs with different mean degrees and matching indices.

\subsection{The Modular Network}
\begin{figure}[h]
\centering
\includegraphics[scale = 0.35,trim=1.6cm 1.6cm 1.6cm 1.6cm, clip=true]{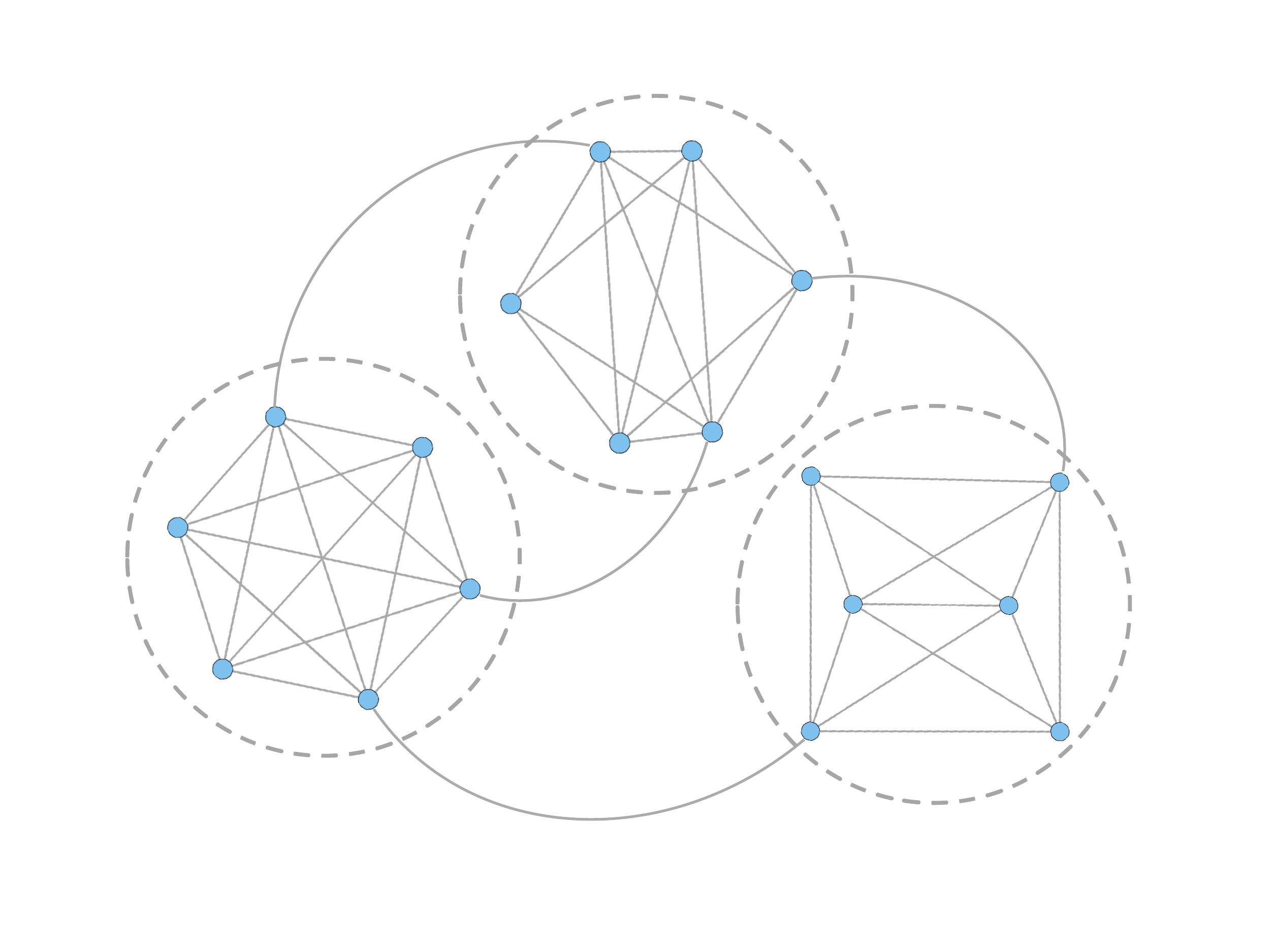}
\caption{Example of a modular network with modules denoted by dotted grey lines.}
\label{fig:clustered_net}
\end{figure}

A subgraph $C$ of a graph $G$ is a set of nodes and edges of $G$ that connect any two nodes in $C$. A \textit{structural module} (or cluster) is rather loosely defined as a highly connected sub-graph with comparatively fewer connections to nodes outside the module~\cite{Fortunato2010}. Conversely, by taking a dynamical perspective,  popularised within the community detection literature, a module (or community) corresponds to a set of nodes and edges were a random walker is likely to become transiently trapped for a longer period of time than that expected at random~\cite{zhou,Delvenne2010,Schaub2012, maur}. A prototypical example of a module would be a complete graph (or clique), where every node is connected to every other node, which is only weakly connected to other nodes. Figure \ref{fig:clustered_net} provides an example of such a modular network. 

To make this notion more precise, we consider the \textit{mismatch index} between a pair of nodes $i$ and $l$, which corresponds to the complement of the matching index defined in Eq.~\eqref{eq:match}:
\begin{equation}\label{miss}
\mu_{il} = |\Gamma(i) \cup \Gamma(l)| - |\Gamma(i) \cap \Gamma(l)| = |\Gamma(i) \cup \Gamma(l)| - I_{il}
%{\sum_{j=1}^N|A_{ij} - A_{lj}| - 2A_{il}} = k_i + k_l - 2(A + A^2)_{il}.
\end{equation}
%\begin{OBS} \label{matchingremark}
%There is a relation between the matching index $I_{il}$ and the mismatch index $\mu_{il}$. 
While the matching index counts all nodes that $i$ and $l$ share, the mismatch index counts all nodes that $i$ and $l$ do not share. 
%More formally
%\begin{eqnarray*}
%\mu_{il} = |\Gamma(i) \cup \Gamma(l)| - |\Gamma(i) \cap \Gamma(l)| = |\Gamma(i) \cup \Gamma(l)| - I_{il}
%\end{eqnarray*}
Hence the normalised mismatch index is:
\begin{align*}
\widehat{\mu}_{il} = \frac{ |\Gamma(i) \cup \Gamma(l)| - |\Gamma(i) \cap \Gamma(l)|}{ |\Gamma(i) \cup \Gamma(l)|} 
%&=& \frac{k_i + k_l - 2(A + A^2)_{il}}{k_i + k_l - (A + A^2)_{il}},\\
= 1 - \widehat{I}_{il}.
\end{align*}
%\end{OBS}
Clearly $\widehat{\mu}_{il} = 0$ if nodes $i$ and $l$ share exactly the same neighbours (as well as potentially being linked themselves), and $\widehat{\mu}_{il} = 1$ if nodes $i$ and $l$ share no common neighbours.

\newpage
\begin{svgraybox}
The theory we present in this chapter holds for all modules $C$ of a network G such that
\begin{equation}\label{relation1}
\frac{\mu_{il}}{{\langle k \rangle}_C} = O \left( \frac{1}{{\langle k \rangle}_C} \right ), ~~ \forall i,l \in C
\end{equation}
along with the homogeneity condition
\begin{equation}\label{relation2}
k_i \approx {\langle k \rangle}_C, ~~ \forall i \in C.
\end{equation}
This final relation also implies that
\begin{equation*}
\frac{\big{|}k_i - k_l \big{|}}{{\langle k \rangle}_C} = O \left( \frac{1}{{\langle k \rangle}_C} \right), ~~ \forall i,l \in C.
\end{equation*}
\end{svgraybox}
Therefore from our first relation (Eq. \eqref{relation1}), a module will have a high matching index, that is, nodes within a module will have a large number of common neighbours. Our second relation (Eq. \eqref{relation2}), allows us to approximate the degree of each node in $C$ by the module mean degree and, certainly for small networks, excludes the possibility of any hubs (highly connected nodes) occurring in a module. As a consequence, module $C$ can be described by the number of nodes it contains $q$ and its mean degree: $C = C(q, {\langle k \rangle}_C)$.

We note that the matching index does not distinguish between a lack of common neighbours within the module and additional (unshared) links to nodes outside the module. Therefore, a high matching index not only guarantees that nodes within a module share similar neighbours but that they also have a comparatively low number of links to nodes outside the module. In \cite{clustsynch}, the edges of a graph were decomposed into inter-module and intra-module edges. They showed that this classification of edges distinguishes the formation of functional modules.

\subsection{The Dynamical Model: Network of Diffusively-Coupled Bounded Systems}\label{sec:diffusive}
We now introduce some dynamics on each node of the graph. The dynamics of each node is governed by $m$-dimensional dynamics:
\begin{equation}
\label{dynamics}
{\frac{d \bm x}{dt} = \bm f(\bm x)},
\end{equation}
where $\bm f: \mathbb{R}^m \rightarrow \mathbb{R}^m$ is a smooth vector field, and we also assume that the solutions of this isolated system are bounded, i.e., for all $t$ there exists a $K$ such that $|| \bm x(t) || < K$. The boundedness of the dynamics of the nodes encompasses a wide variety of stationary and oscillatory (periodic and chaotic) systems \cite{evo}. 

The influence that neighbour $j$ exerts on the dynamics of node $i$ is assumed to depend on the difference of their state vectors: ${\bm x}_{j}(t) - {\bm x}_{i}(t)$. This type of coupling tries to equalise all states of the nodes and it is in this sense that it is called a \textit{diffusive coupling}. The model accounts for the influence of all neighbours in a network $G$ with adjacency matrix $\bm A$, which is assumed to be given. The dynamics of node $i$ in a network of $N$ diffusively coupled elements is then given by:
\begin{equation}\label{md1}
{\frac{d {\bm x}_{i}}{dt} =  \bm f( \bm x_{i}) + \alpha \sum_{j = 1}^{N} A_{ij} \left [ \bm H({\bm x}_{j}) - \bm H({\bm x}_{i}) \right]},  \quad  i=1,\ldots,N
\end{equation}
where $\alpha \in \mathbb{R}$ is the overall coupling strength and $\bm H :  \mathbb{R}^m \rightarrow  \mathbb{R}^m$ is the coupling function. Hence the coupling between elements is given in terms of the adjacency matrix and $\alpha$ modulates the influence between connected nodes. Note that we assume identical elements, i.e., in our model, the dynamics $\bm f$ and coupling $\bm H$ are identical for all nodes. 

\section{Results}\label{results}

In this section, we first state briefly our main analytical results, which are derived in detail in Section~\ref{derivation}. We then provide extensive numerical simulations to illustrate our findings.

\subsection{Summary of Analytical Results} 
Firstly, we state our assumptions regarding the dynamics of the system.
\begin{itemize}
\item[ \textbf{A1}] $\quad$:   The coupled node dynamics in Eq. (\ref{md1}) are bounded: there is a constant $K_x$  such that 
$${\| \bm x_i (t)  \|} < K_x,  \, \,  \, \forall i.$$  

\item[ \textbf{A2}] $\quad$: The variational equation  
\begin{equation}\label{xi}
\dot{\bm{\xi}} = \left [D\bm{f}(\bm{s}(t)) - \sigma D\bm{H}(\bm{s}(t)) \right ] \bm{\xi},
\end{equation}
where $\bm{s}(t)$ is the trajectory of any node,  
admits a uniformly asymptotic trivial solution for
 $\sigma \in (\lambda, \Lambda)$, 
where the both the upper and lower bounds depend on the dynamics and coupling functions: $\lambda = \lambda(\bm f, \bm H)$ and $\Lambda = \Lambda(\bm f, \bm H)$ 
\footnote{
We see the equation as a parametric equation in the same spirit as the master stability function approach \cite{Barahona2002,pec,PRL_2013}, hence we omit the subindex that explicitly shows the dependence on the node.}
. That is, the solution of the variational equation is a contraction:
$$ \bm{\xi}(t) = \bm{T}(t,s) \bm{\xi}(s) \quad \text{for $t \geq s$} $$
with
$$ \| \bm{T}(t,s)  \| \le K e^{-\eta (t-s)} $$
where $\eta = \eta(\sigma) > 0$ uniformly and $K \in \mathbb R$.
\end{itemize}

Assumption A1 is natural in applications. In particular, if a Lyapunov function exists for the isolated dynamics~\eqref{dynamics} with an absorbing domain, it is possible to show that the network solution satisfies A1~ \cite{tiago}. Assumption A2 is similar to the master stability function approach~\cite{pec}. 
The main difference is that in the master stability function, the trajectory $\bm{s}(t)$ corresponds to the a modified (perturbed) solution of the uncoupled dynamics, whereas here it corresponds to the trajectory of a coupled node. 
Depending on the structure of the coupling function this difference is immaterial \cite{Nonlinearity}.  Our numerical analysis shows that the values of $\lambda$ and $\Lambda$ from the master stability function provide a good approximation.

\begin{OBS}\label{positive}
If the coupling function $\bm{H}$ is a positive definite matrix, then the 
results of Ref. \cite{Nonlinearity} demonstrate that $\lambda = \lambda(\bm{f},\bm{H})$ and $\Lambda \rightarrow \infty$. Moreover, 
the contraction exponent is given by 
$$
\eta = \beta \sigma - \lambda
$$   
where $\beta$ is the smallest eigenvalue of $\bm{H}$. 
\end{OBS}

Using these two assumptions in combination with the modular structure of the network, we derive, in Section \ref{derivation}, a stability condition for the synchronisation of modules which does not require a spectral analysis of the network adjacency matrix. This is a consequence of the high matching index within modules. Our results enable the prediction of functional module formation and disintegration depending on the structural properties of the module and the dynamical properties of the nodes. 

Our main finding is the following:
\begin{svgraybox}
{\it 
Consider the modular network $G$ containing a module $C = C(q,{\langle k \rangle}_C)$ with $q \gg 1$ nodes and mean degree ${\langle k \rangle}_C$. Assume that A1 and A2 hold, such that the system in Eq.~\ref{xi} is a contraction for $(\lambda,\Lambda)$. If the matching index of the module ${\langle \widehat{I} \rangle}_C$ is high, then the critical coupling strengths $\alpha^s_C$ for synchronisation  and $\alpha_C^d$ for desynchronisation are given by
\begin{equation}\label{Thm1}
\alpha_C^s = \frac{\lambda}{{\langle k \rangle}_C}(1+ \varepsilon_s) \, \, \, \mbox{   and   } \, \, \,  \alpha_C^d = \frac{\Lambda}{{\langle k \rangle}_C}(1+ \varepsilon_d),
\end{equation}
where $\varepsilon_{s,d} = O(1/{\langle k \rangle}_C)$.
Hence, for $\alpha \in (\alpha_C^s, \alpha_C^d)$, $\exists T \in \mathbb R$ such that $\forall t > T$, the nodes in $C$ exhibit stable synchronised dynamics
\begin{equation}\label{Thm2}
\| \bm x_{i}(t) - \bm x_{l}(t) \| \leq {O}\left( \frac{1}{{\langle k \rangle}_C }\right), \forall i,l \in C.
\end{equation}
}
\end{svgraybox}

This result shows that, under these assumptions, the average degree of the module has an effect on the coupling interval that guarantees synchronisation in the module: only if the coupling $\alpha$ is such that $\alpha_C^s<  \alpha<\alpha_C^d$  the functional and structural modules coincide. If the coupling strength is too large ($\alpha > \alpha_C^d$) or too small ($\alpha < \alpha_C^s$), the functional module disintegrates and no longer reflects the structural module. We call this change the bifurcation between functional and structural modules.  Importantly, the intervals in which synchronisation is stable will be different for different modules, depending on their mean degree.

%\begin{OBS}
%If the coupling function $\bm H$ is a positive definite matrix, then $\alpha_C^d \rightarrow \infty$. Therefore for this class of coupling function, if $\alpha > \alpha_C^s$, the functional module mirrors the structural module, and no bifurcation between structural and functional modules is observed. However, clearly if $\alpha < \alpha_C^s$, the functional modules will not resemble the structural modules. This is the case of the coupling $H=I$ as demonstrated in the Section \ref{num_sims}.
%\end{OBS}
%\MB{This remark was unclear to me... I have repeated it below as I interpreted it but I might be wrong about it... Please check and use/discard from my version at will...}
%\textcolor{blue}{
\begin{OBS}
If the coupling function $\bm H$ is a positive definite matrix, then $\alpha_C^d \rightarrow \infty$. Therefore for this class of coupling function and large enough values of the coupling $\alpha > \max \{\alpha_C^s\}$, the functional modules mirror all the structural modules in the network. On the other hand, if $\alpha < \min \{\alpha_C^s\}$ no functional modules will be apparent. In between those two limits, only some of the structural modules will be reflected as functional modules.
This is the case of the coupling $H=I$.
\end{OBS}
%}

%\clearpage
Note also that the solutions $\bm x_{i}$ under cluster synchronisation may not be similar to the solutions of the invariant synchronisation manifold {$S$} of the whole network
\begin{equation*}
S = \{ \bm x_{i}(t) = \bm s(t) \text{ where } \dot {\bm s} = \bm f(\bm s), \forall i = 1, ... , N \}.
\end{equation*}
Therefore the dynamics of nodes in different modules can be very different to each other and, in particular, to the global synchronous dynamics of the network. 
In our analysis, we effectively decompose a modular network into individual modules with low inter-module connectivity and predict the onset of stable synchronisation based upon the mean degree within a module. For the derivation of our analytical results, see Section \ref{derivation}.

\subsection{Numerical Simulations}\label{num_sims}

To illustrate our analytical results we consider numerical simulations of the paradigmatic example of a network of diffusively coupled identical R\"{o}ssler oscillators. The isolated dynamics of each oscillator $i$ is described by the system of differential equations
\[ \dot{\bm x}_i = {(\dot{x}_i, \dot{y}_i, \dot{z}_i)}^{T} = \bm f(\bm x_i) =
 \left( \begin{array}{cc}
- (y_i + z_i)\\
x_i + ay_i\\
b + z_i(x_i - c) \end{array} \right),\]
with the standard parameter values $a = 0.2, b = 0.2$ and $c= 9$. For these values, we know the system exhibits a chaotic attractor and that all trajectories eventually enter a compact set, thereby satisfying our assumptions regarding $\bm f$ from Section \ref{sec:diffusive}.

For a network of $N$ diffusively coupled R\"{o}ssler oscillators, the dynamics of a node $i$ are governed by the diffusive model (Eq.\eqref{md1}), repeated here for clarity
\begin{equation} \label{diffusive}
\frac{d {\bm x}_{i}}{dt} =  \bm f( \bm x_{i}) + 
\alpha \sum_{j = 1}^{N} A_{ij} \left[ \bm H({\bm x}_{j}) - \bm H({\bm x}_{i})\right],  \quad i=1,\ldots,N,
\end{equation}
where $\alpha \in \mathbb R$ is the global coupling strength, $\bm H \in \mathbb{R}^{3 \times 3}$ is the inner coupling matrix, and $\bm A \in \mathbb R^{N \times N}$ encodes the graph topology.

To numerically determine the stability of synchronisation for a system of coupled oscillators, we construct a correlation matrix $\bm \rho(\alpha) \in \mathbb {R}^{N \times N}$ for a particular coupling strength $\alpha$. This correlation matrix describes the pairwise similarity between the dynamics of all oscillators in the system averaged over some large time $T$ 
\begin{equation*}
\bm \rho(\alpha) = \bm I - \frac{{\bm R}(\alpha)}{\widehat{R}},
\end{equation*}
where the elements of the matrix $\bm{R}(\alpha)$ are defined as:
\begin{equation*}
{R}_{ij}(\alpha) = \frac{1}{T} \sum_{t=0}^{T} \| \bm x_i(t) - \bm x_j(t) \|,
\end{equation*}
for nodes $i$ and $j$ and
\begin{equation*}
\widehat{R} = \max_{i, j, \alpha} {R_{ij}}(\alpha).
\end{equation*}
Using this notation, we also define the mean correlation between the dynamics of a set $S$ of $n$ nodes
\begin{equation*}
{\langle \rho \rangle}_{S} = \sum_{\substack{i,j  \in S \\ i \neq j}} \frac{\rho_{i j}}{n(n - 1)}.
\end{equation*}

We integrated Eq. \eqref{diffusive} using an Adams-Bashforth multi-step scheme together with an intial fourth order Runge-Kutta algorithm using a step size of 0.001.  The initial states of the oscillators were randomised between $0$ and $0.05$. 

We calculated the correlation matrix $\bm \rho$ for a network with a modular structure, varying coupling strengths $\alpha$ and for two coupling schemes: $\bm H = \bm I$ and $\bm H = \bm E$, where
\[ \bm E =  \left( \begin{array}{ccc}
1 & 0 & 0\\
0&0&0\\
0&0&0 \end{array} \right).\]
The case $\bm H = \bm I$ corresponds to each variable $x,y$ and $z$ being coupled to the same variable of all its neighbours, while $\bm H = \bm E$ corresponds to only the $x$ variable being coupled to its neighbouring nodes.

\subsubsection{Dynamics within a modular network}
\begin{figure}[h]
\centering
%\begin{minipage}{1\textwidth}
%\centering
\includegraphics[width=.35\textwidth]{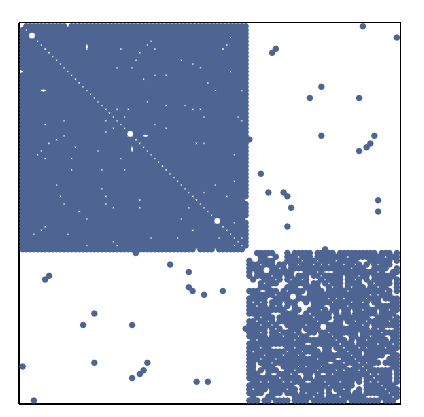}
\includegraphics[trim = 3.5cm 5cm 5cm 0cm clip = true, scale = 0.4,angle = 95]{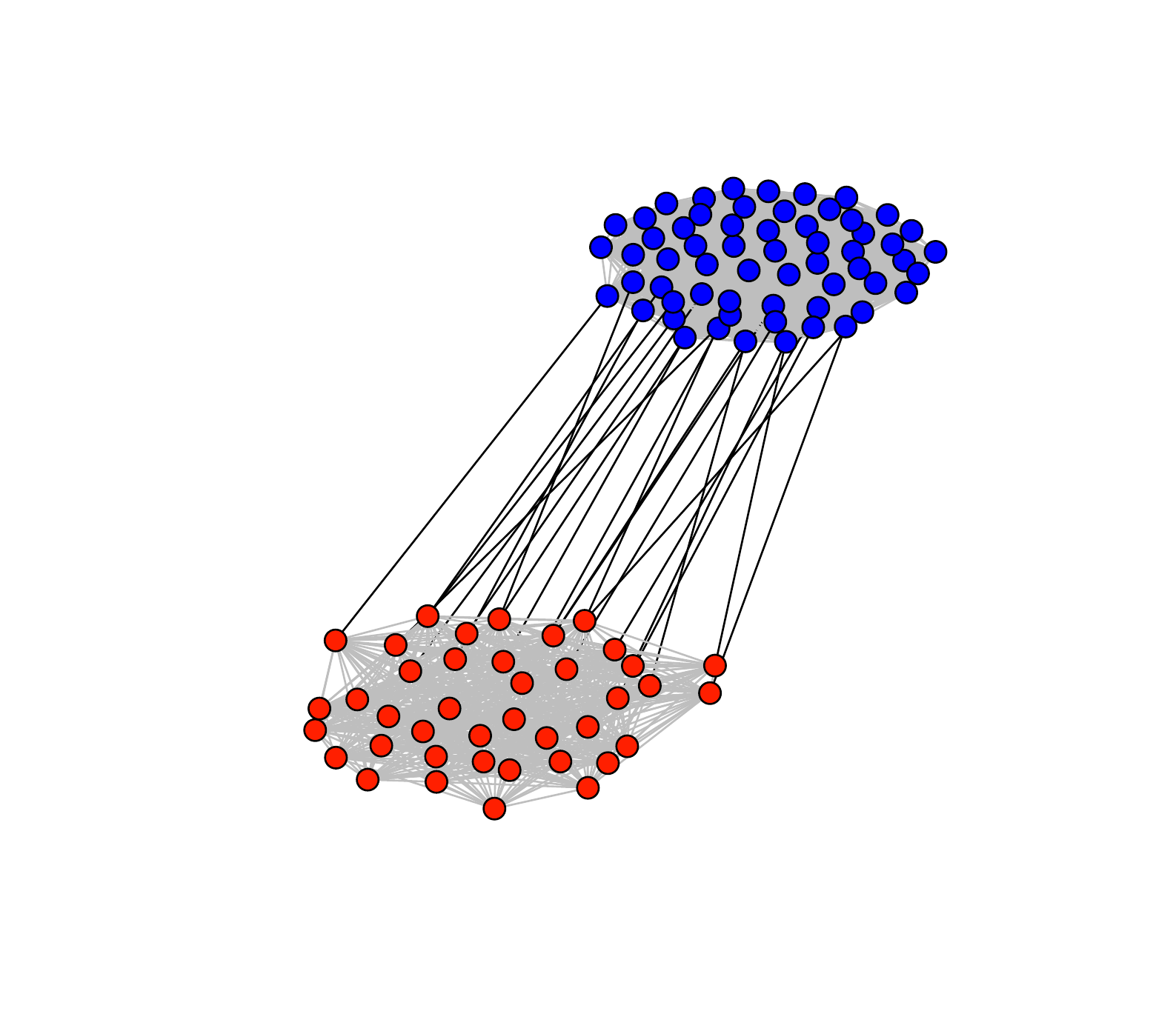}
\caption{Adjacency matrix (\texttt{spy(A)}) and graph visualisation of the network $G_1$ which contains two weakly connected Erd\"os-R\'enyi modules $C_1$ (blue) and $C_2$ (red) with $60$ and $40$ nodes respectively.} 
\label{fig:100_structure}
%\end{minipage}
\end{figure}

\begin{figure}[h]
\centering
 \label{fig:100}
\subfigures
\begin{minipage}{0.32\textwidth}
\caption{\hspace*{.2cm}$\alpha = 0$}
\includegraphics[scale = 0.9]{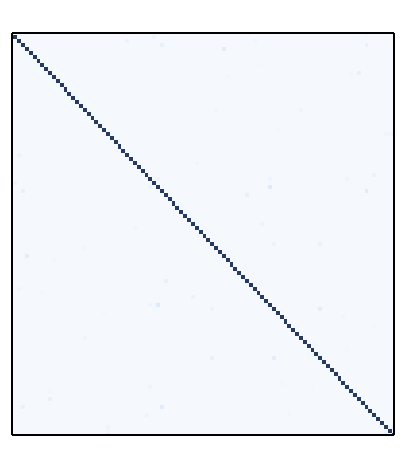}
\label{fig:100_0}%
\end{minipage}
\begin{minipage}{0.32\textwidth}
\caption{\hspace*{.2cm}$\alpha = 3.5 \times 10^{-3}$}
\includegraphics[scale = 0.9]{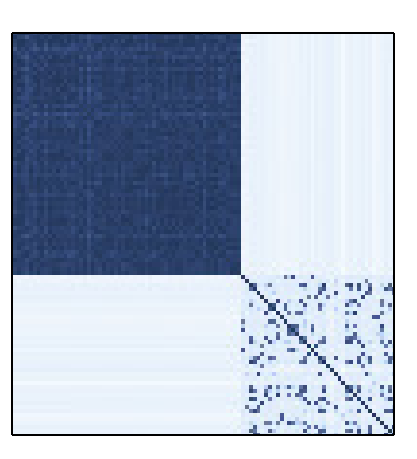}
\label{fig:100_00035}%
\end{minipage}
\begin{minipage}{0.32\textwidth}
\caption{\hspace*{.2cm}$\alpha = 2 \times 10^{-2}$}
\includegraphics[scale = 0.9]{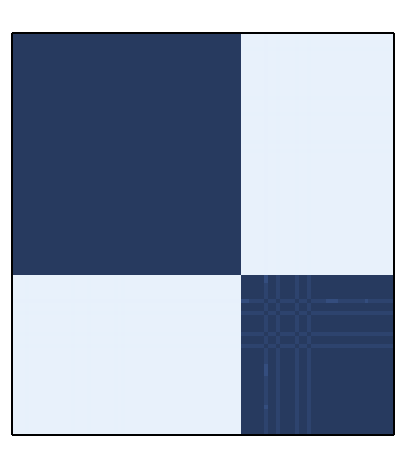}
\label{fig:100_02}%
\end{minipage}
\begin{minipage}{0.32\textwidth}
\caption{\hspace*{.2cm}$\alpha = 1 \times 10^{-1}$}
\includegraphics[scale = 0.9]{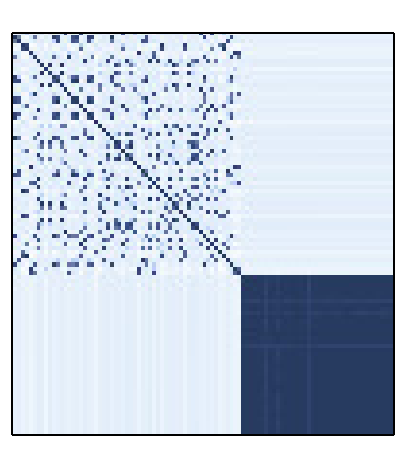}
\label{fig:100_01}
\end{minipage}
\begin{minipage}{0.32\textwidth}
\caption{\hspace*{.2cm}$\alpha = 1.8 \times 10^{-1}$}
\includegraphics[scale = 0.9]{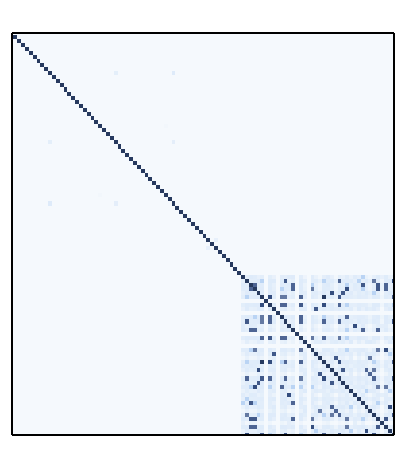}
\label{fig:100_18}
\end{minipage}
\begin{minipage}{0.32\textwidth}
\centering
\includegraphics[trim = 0.2cm 0cm 0cm 0cm, clip = true, scale = 0.28]{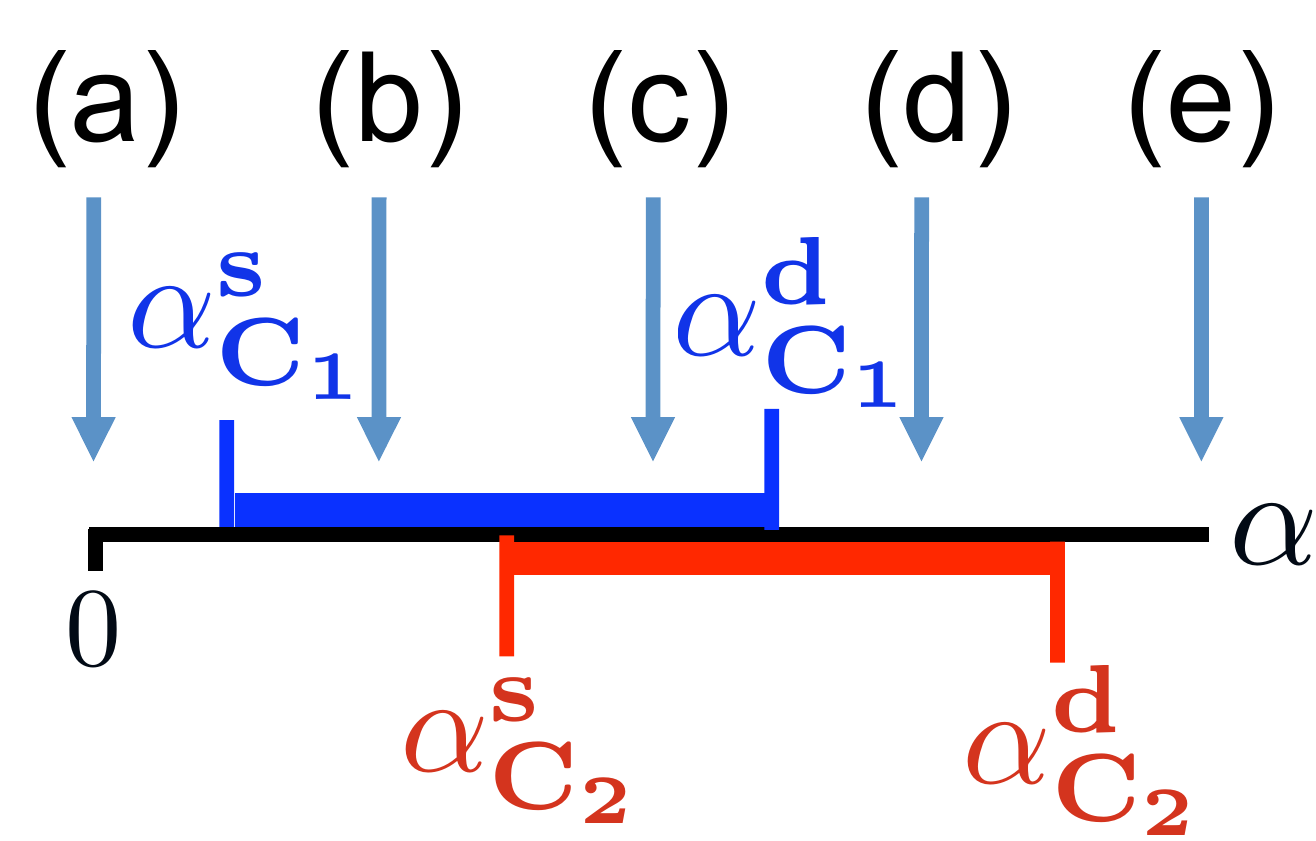}
\label{fig:regions}
\end{minipage}
\samenumber
\caption{Figures \ref{fig:100_0} to \ref{fig:100_18} display the functional correlation matrix $\bm \rho$ for increasing coupling strengths $\alpha$ of the network $G_1$ when only the $x$ component of the dynamics is coupled ($\bm H = \bm E$). Darker coloured areas correspond to regions of synchronisation. The inset illustrates the different synchronisation regions for modules $C_1$ and $C_2$ as expected from our analysis. The letter labellings correspond to those in Figures \ref{fig:100_0} to \ref{fig:100_18}. Note that $\alpha_{C_1}^s \approx 3.2 \times 10^{-3} < \alpha_{C_2}^s \approx 6.2 \times 10^{-3} < \alpha_{C_1}^d \approx 7.2 \times 10^{-2} < \alpha_{C_2}^d \approx 1.39 \times 10^{-1}$, thus giving rise to distinct regions for the functional network.}
\end{figure}

We first present the numerical results of the simulations of the modular network $G_1$ shown in 
Fig.~\ref{fig:100_structure}. The network has two modules $C_1$ and $C_2$ generated according to an Erd\"os-R\'enyi architecture: $C_1$ contains $60$ nodes with ${\langle \widehat{I} \rangle}_{C_1} = 0.95$ and ${\langle k \rangle}_{C_1} = 58$;  $C_2$ contains $40$ nodes with ${\langle \widehat{I} \rangle}_{C_2} = 0.6$ and ${\langle k \rangle}_{C_2} = 30$. The inter-module connections are low compared with the intra-module links, as implied by the high module matching indices. In these simulations, we use the $x$-coupling: $\bm H = \bm E$. The results of the functional analysis are presented in Figure \ref{fig:regions}.

Figures \ref{fig:100_0} to \ref{fig:100_18} display heat map representations of the correlation matrix $\bm \rho$ for increasing coupling strengths $\alpha$. Darker regions correspond to higher correlation between node dynamics. From the analysis of the master stability function of the 
R\"{o}ssler system with these parameters and $x$-coupling,  it has been found that the
region of stable synchronisation are bounded by $\lambda = 0.186$ and $\Lambda = 4.164$ \cite{msf}.  We can then use these numbers to approximate the regions of stable synchronisation for modules $C_1$ and $C_2$:  $\alpha_{C_1}^s \approx 0.186/58 = 3.2 \times 10^{-3}$ and $\alpha_{C_1}^d \approx 4.164/58 = 7.2 \times 10^{-2}$ for $C_1$; whereas $\alpha_{C_2}^s \approx 6.2 \times 10^{-3}$ and $\alpha_{C_2}^d \approx 1.39 \times 10^{-1}$ for $C_2$. These regions are indicated by the illustration inset in Figure \ref{fig:regions}.

Our numerics show that when $\alpha = 0$ there is no correlation between the dynamics of the nodes. This is expected since there is no interaction between the oscillators. For $\alpha = 3.5 \times 10^{-3}$, $C_1$ has synchronised and some nodes in $C_2$ are beginning to show cohesive dynamics. This is expected since the coupling strength has entered the predicted stable synchronisation region for $C_1$ but not for $C_2$. For $\alpha = 2 \times 10^{-2}$, both modules have synchronised, but the dynamics of the two modules are \textit{uncorrelated} as indicated by the pale off-diagonal regions. This is a result of the low inter-module connectivity\footnote{Under certain conditions, it is possible for two modules to synchronise and this is explored elsewhere such as \cite{clustersynch}.}.
For $\alpha = 1 \times 10^{-1}$ (Figure \ref{fig:100_01}), module $C_1$ has already desynchronised, while $C_2$ remains synchronised.  As the coupling strength is increased further, $C_2$ also desynchronises.

\begin{figure}[h]
\centering
\subfigures
\centering
  \caption{\hspace*{.1in} $\alpha = 0$}
    \includegraphics{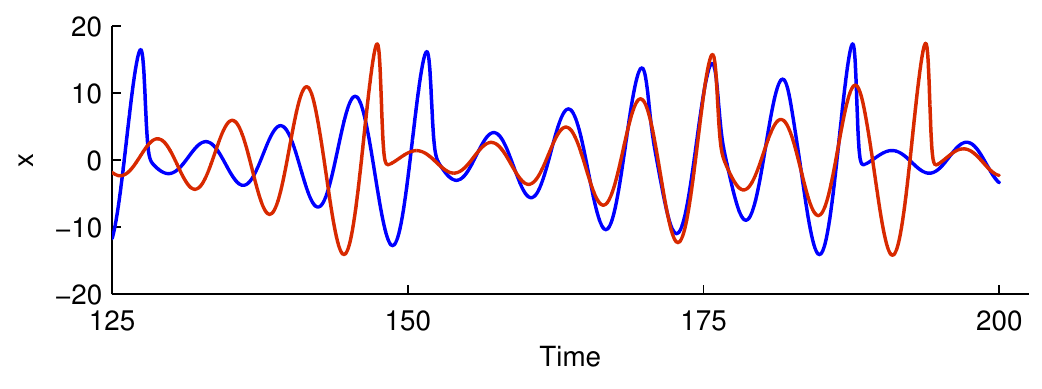}
  \label{fig:a_0} 
  
\caption{ \hspace*{.1in} $\alpha = 3.5 \times 10^{-3}$}
  \includegraphics{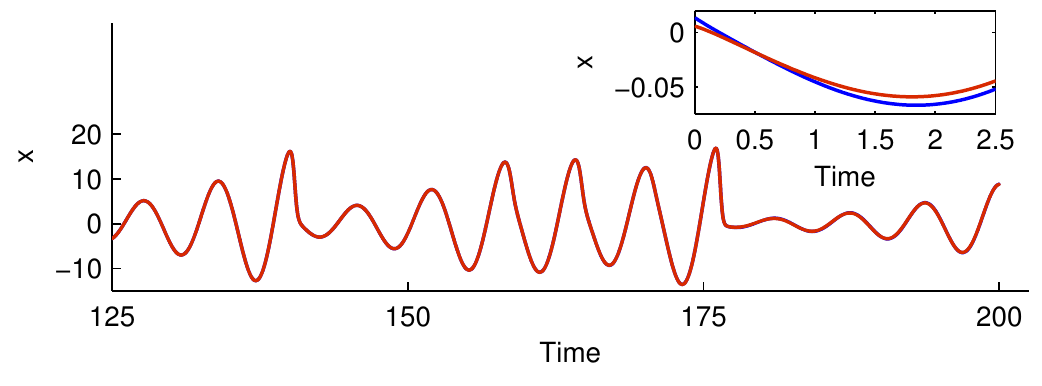}
 \label{fig:a_0_0035}
\centering
\caption{\hspace*{.1in} $\alpha = 1 \times 10^{-1}$}
  \includegraphics{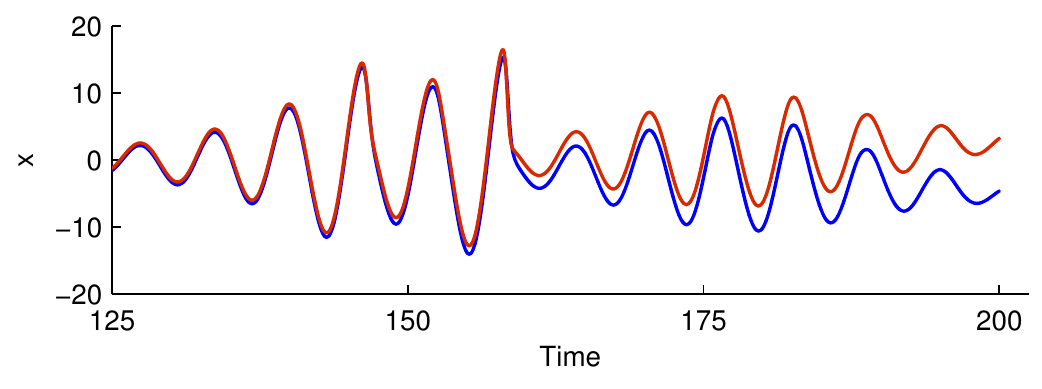}
 \label{fig:a_0_1} 
\end{figure}

The dynamical behaviour of the clusters is perhaps expressed more clearly when analysing the time evolution of the oscillators. Figure \ref{fig:ts} displays the evolution of the $x$ variable of two oscillators in module $C_1$ for increasing coupling strengths. 
%$\alpha=0, 3.5 \times 10^{-3}$ and $1 \times 10^{-1}$. 
In Figure \ref{fig:a_0}, where $\alpha = 0$, we observe, as expected, that the dynamics of the two oscillators are uncorrelated. For $\alpha = 3.5 \times 10^{-3}$ in Figure \ref{fig:a_0_0035}, the two oscillators synchronise. The inset in this figure demonstrates that the two oscillators have differing but close initial conditions and due to the coupling strength, soon achieve stable synchronisation. Finally, for $\alpha = 1 \times 10^{-1}$ (Figure \ref{fig:a_0_1}), the oscillators initially synchronise before diverging after a short time and then remain uncorrelated thereafter.

\subsubsection{Relation between the critical coupling strength and the module mean degree}
We now examine how the critical coupling strength required for stable synchronisation within a module depends upon its mean degree. To this end we simulated a network $G_2$ of $100$ nodes for varying matching indices between $0.7$ and $1$ which will act as a paradigmatic example of a module with no inter-module links, thereby removing any external perturbations from other modules. We then determined the coupling strength $\alpha$ such that ${\langle \rho \rangle}_{G_2} =  0.99$ for both $\bm H = \bm I$ and $\bm H = \bm E$. We repeated simulations $5$ times for varying network adjacency matrices.

Along with the simulations, we indicate the predicted critical coupling strength which depends inversely upon the mean degree of the module as determined by our methodology. 
We demonstrate this by fitting the simulation results to an expression of the form
\begin{eqnarray}\label{fitting}
\alpha_C^s, \alpha_C^d &=& \frac{a}{{\langle k \rangle} + b},
\end{eqnarray}
where $a$ and $b$ are fitting parameters.

\vspace{-3.5em}
\begin{figure}
\samenumber
\caption{Time evolution of the $x$ component of the dynamics for two oscillators in module $C_1$ over the time window $125$ to $200$. For $\alpha = 0$, the dynamics of the two oscillators are uncorrelated. When the coupling strength is increased to $3.5 \times 10^{-3}$, the two oscillators enter a stable synchronised state. As the coupling strength is increased further to $1 \times 10^{-1}$, the oscillators initially synchronise for small times before their trajectories diverge and then remain uncorrelated thereafter.}
\label{fig:ts}
\end{figure}

\clearpage

\begin{OBS}
The precise bounds for the critical coupling strength depends on the actual node degrees and not only the mean degree (see the derivation of the results in Section~\ref{derivation} for details). However, since node degrees are close to the mean degree we can approximate $k_i = \langle k \rangle + b$, where $b$ is treated as a free parameter. The parameter $b$ effectively allows for inhomogeneities within the module structure to produce a small perturbation to the critical coupling strength required for stable synchronisation, details on these perturbations can be found in Section \ref{sec:bounds}.
\end{OBS}

\begin{figure}[h]
\centering
\includegraphics[scale = 0.8, trim = 1cm 0cm 0cm 0cm, clip = true]{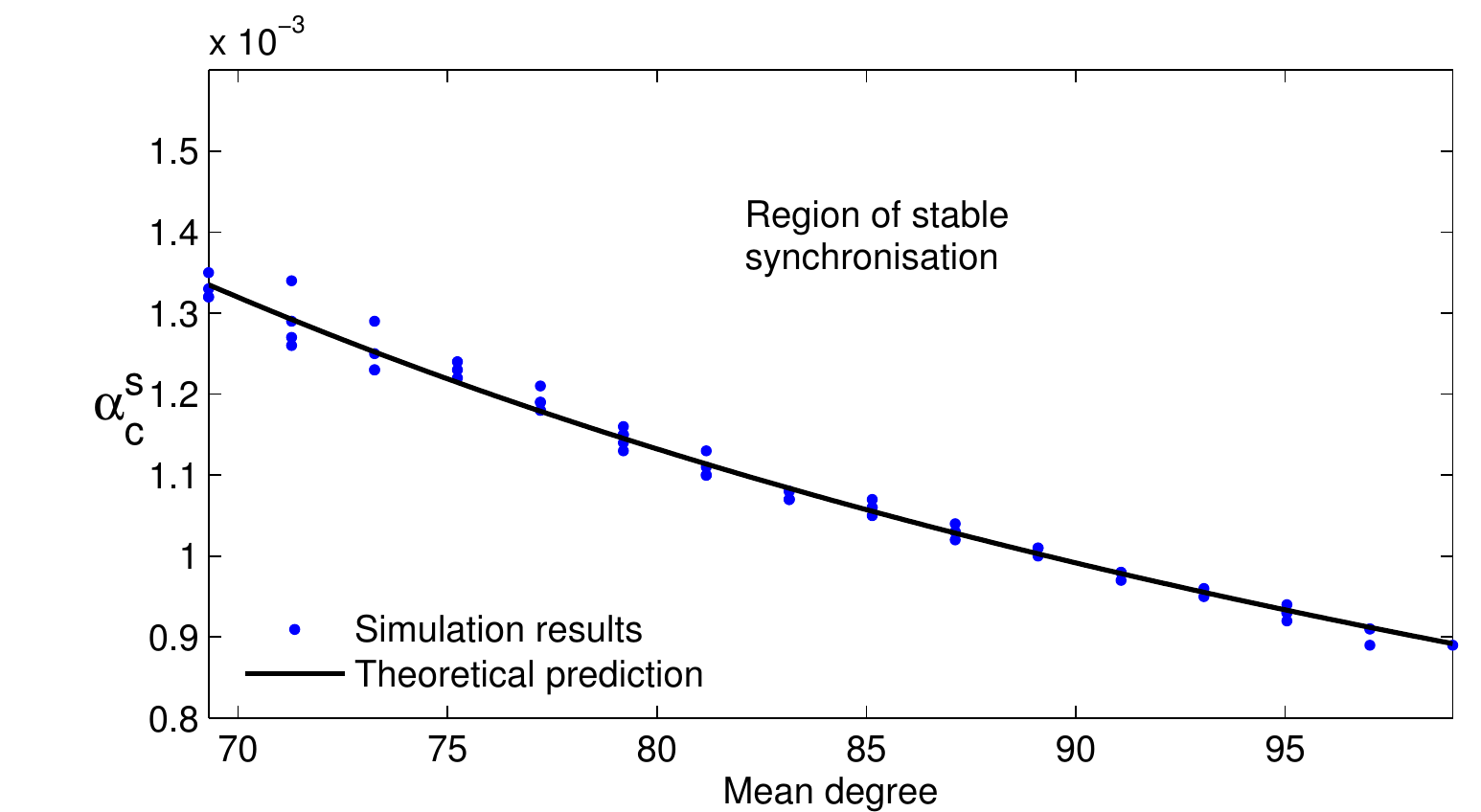}
\caption{Critical coupling strength required for stable synchronisation for a module of $100$ nodes with a varying high matching index plotted against the module mean degree. Simulations were repeated $5$ times with the coupling $\bm H = \bm I$. We also display the fitting curve (Eq. \eqref{fitting}), which depends inversely upon the module mean degree.}
\label{fig:100_vary_all}
\end{figure}

Figure \ref{fig:100_vary_all} displays the results when coupling all components of the dynamics with $\bm H = \bm I$. As expected, all nodes in the module will synchronise given a strong enough coupling strength and the module will remain synchronised as the coupling strength is increased thereafter. The simulations follow an inverse dependence upon the mean degree of the module (Eq. \eqref{fitting}), as expected from our results.

\begin{figure}[H]
\centering
\includegraphics[scale = 0.8,trim = 1.1cm 0cm 0cm 0cm,clip=true]{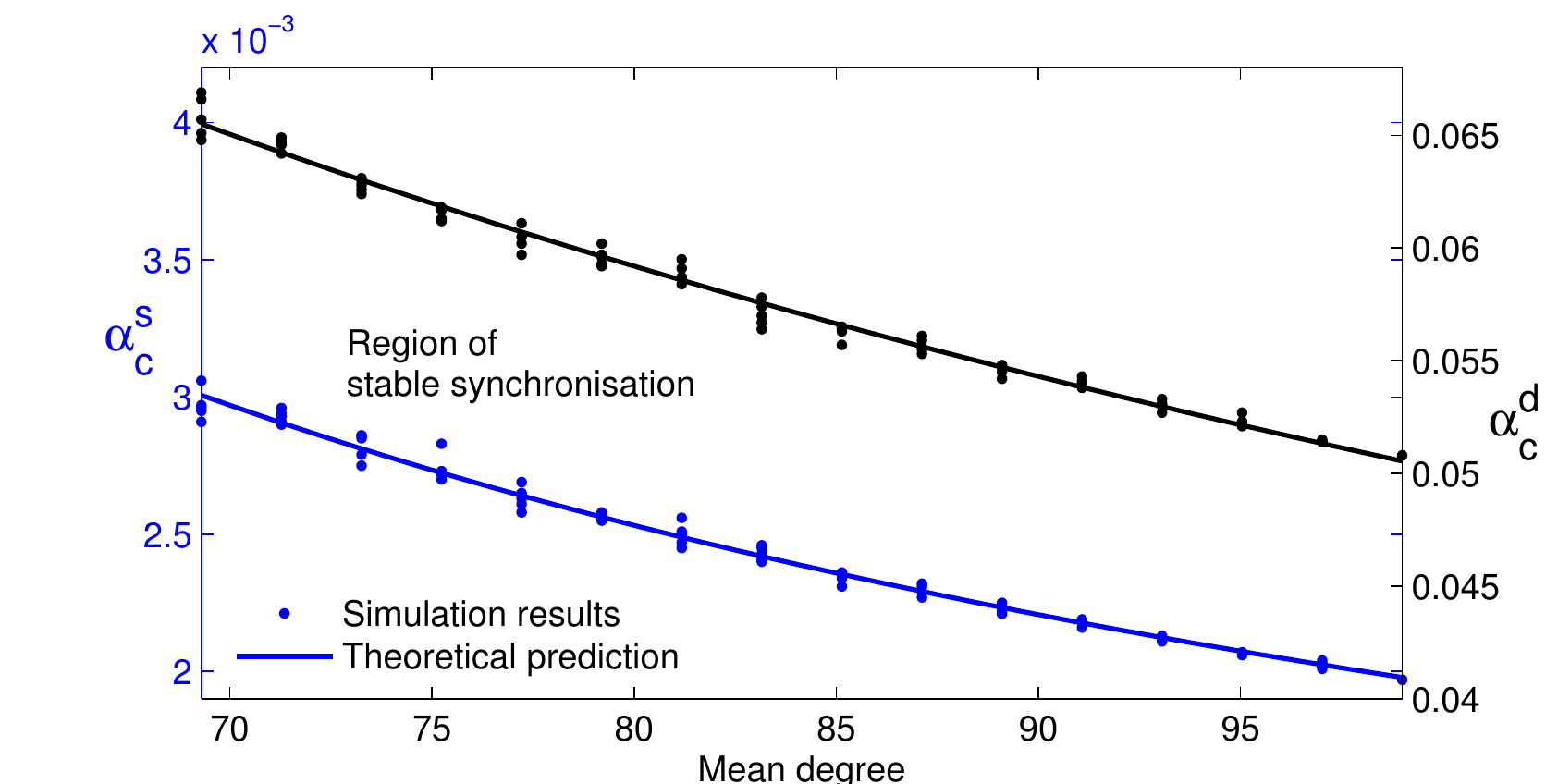}
\caption{Critical coupling strengths required for stable synchronisation of a module of $100$ nodes with varying high matching index plotted against the module mean degree. Simulations were repeated $5$ times with the coupling function $\bm H = \bm E$. We also display the fitting curves, (Eq. \eqref{fitting}) which depend inversely upon the module mean degree. Note the different scales on the y-axes, with blue corresponding to synchronisation and black to desynchronisation.}
\label{fig:100_vary_x}
\end{figure}

Figure \ref{fig:100_vary_x} displays the results when coupling only the $x$ component of the dynamics, corresponding to $\bm H = \bm E$. We see that for a module with a high matching index, all nodes will synchronise above a coupling strength and will then desynchronise as the coupling strength is increased further. Again, from the fitted curves, the critical coupling strengths required for synchronisation and desynchronisation display an inverse dependence upon the module mean degree in line with our results.

%The observed variability in the results is due to fluctuations in the random realisations of the graph topology as well as originating from the finite time computation of the correlations. Our approach provides an expected bound  and we do not consider here this small random variability. The curves serve to provide a validation for our analytical results and the small variations seen between simulations is additionally encouraging.

\begin{OBS}
When setting $b = 0$ in the curve fitting (corresponding to perfectly homogeneous node degrees), we can directly compare our results with those obtained in the literature from analysing the master stability function, which have shown  $\lambda = 0.186$ and $\Lambda = 4.614$ for this system \cite{msf}.  From our fits, allowing for a stable synchronisation region to be given by ${\langle \rho \rangle}_{G_2} > 0.99$, we obtain $\alpha_C^s \langle k \rangle = 0.202$ and $\alpha_C^d \langle k \rangle = 4.750$ in close agreement with the master stability function bounds, as expected from our analytical results.
\end{OBS}

\clearpage

\section{Derivation of our Analytical Results}\label{derivation}

Before detailing the derivation of our results below, we briefly outline our strategy. 
Starting from Eq. \eqref{md1}, we define
\begin{equation}\label{z}
\bm z(t) = \bm x_{i}(t) - \bm x_{l}(t), \mbox{ for any $i,l \in C$,} 
\end{equation}
where $C$ is a module, and analyse the dynamics of $\bm z$. To this end, we take the following four steps.
\begin{enumerate}
\item First, for $\bm z$ sufficiently small, we obtain the linearised equation for $\bm z(t)$:
\begin{equation}\label{pert1}
\frac{d\bm z}{dt} = \bm h(\alpha,t)\bm z(t) + \alpha{\bm g(t)},
\end{equation}
where $\bm{g}$ and $\bm h$ are to be determined and depend on both the node dynamics and network structure. By Taylor's Theorem, the remainder is $O(||\bm z(t)||^2)$ and it can be dealt with in step 2.
% below we guarantee that these higher order perturbations will be damped out. 
\item We then analyse the associated homogeneous equation
\begin{equation*}
\frac{d\bm y}{dt} = \bm h(\alpha,t)\bm y,
\end{equation*}
and, by Eq. \eqref{xi}, represent $\bm y(t)$ in terms of its associated evolution operator
\begin{equation*}
\bm y(t) = \bm T(t,s) \bm y(s).
\end{equation*}
Our assumption A2 guarantees that the trivial solution of the above equation is uniformly asymptotically stable, that is, for some $\eta > 0$,
\begin{equation*}
\|\bm y(t)\| \leq Ke^{-\eta(t-s)}\|\bm y(s)\|, \mbox{  for   } t \geq s.
\end{equation*}
\item We then solve Eq. \eqref{pert1} using the method of variation of parameters 
\begin{equation*}
\bm z(t) = \bm T(t,s) \bm z(s) + \alpha{\int_s^t \! \bm T(t,u) \bm g(u) \, \mathrm{d} u},
\end{equation*}
and by defining $\| \bm g \| = \sup_u \| \bm g (u)\| $, from the triangle inequality we obtain
\begin{equation*}
\| \bm z(t) \| \leq Ke^{-\eta(t-s)} \| \bm z(s) \| + \frac{K \alpha \| \bm g \|}{\eta}.
\end{equation*}
\newpage
\item Then for large times and using Eq. \eqref{z}, we obtain
$$
\| \bm{x}_i - \bm{x}_l  \| \le  \frac{\tilde K \alpha \| \bm g \|}{\eta},
$$
where
$$
\tilde K = K \left( \frac{\eta e^{- \eta (t-s)} || \bm z(s) ||}{\alpha || \bm g ||} + 1\right).
$$
Under our network assumptions, Eq. \eqref{relation1} and Eq. \eqref{relation2}, we can obtain bounds for $\| \bm g \|$ as 
$$
\frac{\| \bm g \|}{{\langle k \rangle}_C} = O \left(\frac{1}{{\langle k \rangle}_C} \right).
$$
\end{enumerate}
We now explain these steps in more detail.

\subsection{Obtaining the Variational Equation}\label{variation}
To obtain the first variational equation for $\bm z(t) = \bm{x}_i(t) - \bm{x}_l(t)$ we write
\begin{eqnarray}
\dot{\bm z} &=& {\dot{\bm x_i}} -  {\dot{\bm x_l}} \nonumber \\ 
%&=& \bm f( \bm x_{i}) - \bm f(\bm x_{l}) +  \alpha{\sum_{j}{A_{ij}}[\bm{H}(\bm x_{j}) - \bm{H}(\bm x_{i})]} -  \alpha{\sum_{j}{A_{lj}}[\bm{H}(\bm x_{j})  - \bm{H}(\bm x_{l})]} \\
&=& \bm f(\bm z + \bm x_{l}) - \bm f(\bm x_{l}) +  \alpha \left\{ {\sum_{j}{(A_{ij} - A_{lj})\bm{H}(\bm x_{j}}}) +   {\sum_{j}{[ A_{lj} \bm{H}(\bm x_{l}) - A_{ij}\bm{H}(\bm x_{i}) ]}} \right\} \nonumber ,
\end{eqnarray}
by Eq. \eqref{md1}.

For some $t$ such that $\|\bm z(t)\|$ is sufficiently small, we can expand as a Taylor series, and after some manipulations we obtain
\begin{eqnarray}
\dot{\bm z} 
&=& D\bm f(\bm x_{l})\bm z  - \alpha{A_{il} D \bm{H}( \bm z}) + \alpha \left\{k_{l} \bm{H}(\bm x_{l}) - k_{i} \bm{H}(\bm x_{i}) +  {\sum_{j\neq{i,l}}{(A_{ij} - A_{lj})\bm{H}(\bm x_{j})}} \right\} \nonumber
\end{eqnarray}
where $k_{i}$ is the degree of node $i$ as given in Section \ref{defn}. Without loss of generality, we assume  $k_{l} \geq k_{i}$ and set $\bar{k}_i = k_i + A_{il}$ to obtain 
\begin{eqnarray}\label{all}
\dot{\bm z}(t) &=& \bm{h}(\alpha, t) \bm z + \alpha \bm{g}(t),
\end{eqnarray}
where 
\begin{equation}
\bm{h}(\alpha, t) = D \bm f(\bm x_{l}(t)) - \alpha \bar k_{i} D \bm{H}
\end{equation}
and 
\begin{equation}\label{functiong}
\bm{g}(t) = { {(k_{l} - k_{i})\bm{H}(\bm x_{l}(t))  + \sum_{j\neq{i,l}}{(A_{ij} - A_{lj}) \bm{H}(\bm x_{j}(t))}}}.
\end{equation}

This is the first variational equation. Note, we truncated our Taylor expansion in $\bm z$ up to first order and by Taylor's Theorem we know the remainder satisfies $\|\bm R(\bm z(t))\| = O(\|\bm z(t)\|^2)$.

\subsection{The Homogeneous Equation}

We now consider the homogeneous part of Eq. \eqref{all}
\begin{eqnarray}\label{hom}
\dot{\bm y} &=& \bm{h}(\alpha, t) \bm y.
\end{eqnarray}
Notice the rescaling $\alpha \bar{k}_i = \sigma$ brings the above equation (Eq. \eqref{hom}) to Eq. \eqref{xi}. Therefore, 
if 
$$
\lambda < \alpha \bar{k}_i < \Lambda,
$$
Eq. \eqref{hom} has an evolution operator satisfying 
\begin{equation}\label{evolution}
\| T(t,s) \| \le K e^{-\eta (t-s)}.
\end{equation}

Now, since modules will have a large number connections and nodes will share many common neighbours, $\bar k_{i}$ will be close to the mean degree ${\langle k \rangle}_C $  (see our network assumptions Eq. \eqref{relation1} and Eq. \eqref{relation2}). This means that in leading order in ${\langle k \rangle}_C$, the stability condition is given by 
$$
\lambda < \alpha \left({\langle k \rangle}_C + \delta \right) < \Lambda,
$$
where $|\delta| = o({\langle k \rangle}_C)$ takes into account the fluctuation between $k_i$ and the mean degree ${\langle k \rangle}_C$. After some rearrangements we obtain
\begin{equation}\label{stability}
\frac{\lambda}{{\langle k \rangle}_C} \left(1 - \frac{\delta}{{\langle k \rangle}_C} \right) < \alpha < \frac{\Lambda}{{\langle k \rangle}_C} \left(1 -  \frac {\delta}{{\langle k \rangle}_C} \right).
\end{equation}
Then for ${\langle k \rangle}_C$ large we see that Eq. \eqref{stability} resembles Eq. \eqref{Thm1}.
Furthermore, according to Remark \ref{positive}, if $\bm{H}$ is positive definite, we obtain $\Lambda \rightarrow \infty$ and 
\begin{equation}\label{stabeta}
\eta = \alpha \beta ({\langle k \rangle}_C + \delta) - \lambda.
\end{equation}
Then for $\alpha$ satisfying $\eta >0$ in Eq. \eqref{stabeta}, we obtain a uniform contraction.
%Therefore the trajectories of the oscillators in $C$ converge over time and their rate of convergence is given my the exponent $\eta$ in Eq. \eqref{stabeta}.
It remains to show that the perturbations will not destroy the synchronisation property.

\subsection{The Perturbed Equation}
We now turn our attention to the inhomogeneous Eq. \eqref{all}. Using the method of variation of parameters we obtain
\begin{eqnarray}
\bm z(t) &=& \bm T(t,s) \bm z(s) + \alpha{\int_s^t \! \bm T(t,u) \bm g(u) \, \mathrm{d} u}, \nonumber
\end{eqnarray}
and by virtue of the triangle inequality we find
$$
\| \bm z(t) \| \leq  \| \bm T(t,s)\| \,  \| \bm z(s)\| + \alpha{\int_s^t \! \| \bm T(t,u) \| \,  \| \bm g(u) \| \, \mathrm{d} u}.
$$
Then, using the bounds for the evolution operator from Eq. \eqref{evolution} we obtain
\begin{eqnarray}
\| \bm z(t)\| &\leq& Ke^{-\eta(t-s)}\| \bm z(s)\| + {\alpha}{\int_s^t \! Ke^{-\eta(t-u)}\| g \| \, \mathrm{d} u}  \nonumber \\
&=& Ke^{-\eta(t-s)}\| \bm z(s)\| + {\alpha} K\| g \| \left[ {\frac{1 - e^{-{\eta(t-s)}}}{\eta}} \right]. \nonumber
\end{eqnarray}
For $t$ large we obtain
\begin{equation*}
\| \bm z(t)\| \leq  {\frac{\tilde K \alpha \| \bm{g} \|}{\eta}},
\end{equation*}
where
\begin{eqnarray*}
\tilde K &=& K \left( \frac{\eta e^{- \eta (t-s)} || \bm z(s) ||}{\alpha || \bm g ||} + 1\right) \\
&=& K + o(1).
\end{eqnarray*}

%As $t \rightarrow \infty$ we obtain
%\begin{equation*}
%\| \bm z(t)\| \leq  {\frac{K \alpha \| \bm{g} \|}{\eta}}.
%\end{equation*}

Now, under the stability condition Eq. \eqref{stability} for $\alpha$, we obtain
\begin{equation}\label{zbound}
\| \bm z(t)\| \leq  {\frac{\tilde {K} \Lambda \| \bm{g} \|}{{\langle k \rangle}_C \eta}}.
\end{equation}
If $\bm H$ is positive definite then from Remark \ref{positive}
\begin{eqnarray}\label{z1b}
\| \bm z(t)\| &\leq&  {\frac{\tilde K \alpha \| \bm{g} \|}{\alpha \beta {\langle k \rangle}_C - \lambda}}.
\end{eqnarray}
We must now analyse the bounds for $\bm g$ within a module.

\subsection{Bounds for the Perturbation} \label{sec:bounds}
We give the argument for Eq. (\ref{zbound}). The argument for Eq. (\ref{z1b}) is similar.  We first recall the mismatch index Eq. \eqref{miss}:
$$
\mu_{il}=  \sum_{j = 1}^N |A_{ij} - A_{lj}| - 2A_{il}. \nonumber
$$
Then since since the trajectories are bounded and the coupling function is smooth, we can bound
$$
\| \bm{H}(\bm{x}_i) \| \le K_h,
$$
for a positive constant $K_h$, and since
$$
\bm{g}(t) = { {(k_{l} - k_{i})\bm{H}(\bm x_{l}(t))  + \sum_{j\neq{i,l}}{(A_{ij} - A_{lj}) \bm{H}(\bm x_{j}(t))}}},  \nonumber
$$
from Eq. \eqref{functiong}, we obtain
\begin{equation}\label{pert}
\frac{\|\bm{g}(t)\|}{{\langle k \rangle}_C} \le \frac{1}{{\langle k \rangle}_C}{ K_h (k_{l} - k_{i} + \mu_{il}) }.
\end{equation}

Then, motivated by the matching index notation in Eq.~\ref{eq:match}, we can introduce 
$$
K_1 = \frac{K_h}{{\langle k \rangle}_C} (k_i + k_l - (A + A^2)_{il})
$$
to obtain
\begin{equation}\label{g_condition}
\frac{\|\bm{g}(t)\|}{{\langle k \rangle}_C} \le {  \frac{|k_{l} - k_{i}|}{{\langle k \rangle}_C}  K_h+ (1 - \widehat{I}_{il}) K_1 }.  \end{equation}
Therefore, for the perturbation to be small, we require:
\begin{itemize}
\item The difference in node degrees within a module to be low.
\item Nodes to share many common neighbours within a module.
\item Nodes to have comparatively fewer connections to nodes outside the module.
\end{itemize}
The final two requirements emerge via the matching index in Eq. \eqref{g_condition} since $\widehat{I}_{il}$ penalises for not only a lack of shared nodes within a module but also for additional (unshared) connections to nodes outside the module.
Then, from our network assumptions (Eq. \eqref{relation1} and Eq. \eqref{relation2}) along with $\alpha$ statisfying the stability condition Eq. \eqref{stability}, by Eq. \eqref{zbound} and Eq. \eqref{pert}, we obtain
$$
\| \bm z(t) \| \le O \left(\frac{1}{{\langle k \rangle}_C} \right).
$$

Therefore the stability of module synchronisation depends upon the module mean degree and the extent to which synchronisation can be achieved depends upon the matching index within a module. 
These conditions will clearly vary between modules and our conditions only guarantee individual modules to synchronise independently as opposed to global network synchronisation.

\section{Conclusion and Discussion}\label{summary}

We have shown that the stability of module synchronisation in complex modular networks can be predicted based upon the module mean degree given certain assumptions on the component dynamics and network structure. Our key assumption on the modular structure is that nodes within modules share many common neighbours and inter-module connections are weak in comparison.

Our analysis revealed two basic scenarios for module synchronisation. If the coupling function is linear and positive definite, we showed that the functional modules reflect the structural modules. In this case, as the coupling strength is increased, the module with the largest mean degree synchronizes first, then more and more modules achieve synchronisation. In this case, the dynamics of the network mirror the structural properties of the network. 

However, for more general couplings, typically our stability criterion A2 is satisfied, see Ref. \cite{msf}. In this case, we observe interesting dynamical behaviour as the coupling parameter is increased where, in a first stage, modules of synchronised nodes can form and reflect the structural organization, but for large couplings the synchronisation becomes unstable and functional modules disintegrate. This scenario corresponds to bifurcations between the functional and structural properties.

Our assumptions on the structure of the modules allowed for an analytical treatment of these scenarios and enabled us to determine the critical coupling strengths for synchronisation and desynchronisation. Additionally, we showed that the module matching index dictates the quality of synchronisation. Moreover, our present approach can be used to explain the synchronised dynamics of groups of nodes not necessarily forming modules such as hub synchronization \cite{Murilo,HubPRE}, since their matching index may be high and the degrees similar.

These results can be of importance for functional network analysis and enhance our understanding of the relation between the functional and structural network. Indeed, even though a module may possess the structural properties required for synchronisation, the functional modules may not reflect the structural modules of a network.

%%%%%%%%%%%%%%%%%%%%%%%% referenc.tex %%%%%%%%%%%%%%%%%%%%%%%%%%%%%%
% sample references
% %
% Use this file as a template for your own input.
%
%%%%%%%%%%%%%%%%%%%%%%%% Springer-Verlag %%%%%%%%%%%%%%%%%%%%%%%%%%
%
% BibTeX users please use
% \bibliographystyle{}
% \bibliography{}
%

\end{document}